\documentclass[11pt]{article}

\usepackage{amsmath}
\usepackage{amsfonts}
\usepackage{amsthm}
\usepackage{amssymb}
\usepackage{amscd}
\usepackage[british]{babel}
\usepackage{graphicx}
\usepackage{psfrag}
\usepackage{epsfig}
\usepackage{rotating}
\usepackage{times}
\usepackage{color}
\usepackage{float}
\usepackage{ulem}

\theoremstyle{plain}

\newcommand{\boxend}{\flushright{$\Box$}}

\begin{document}



\title{Reheating  via gravitational particle production in simple models of quintessence or $\Lambda$CDM inflation}

\author{
Jaume Haro$^{a,}$\footnote{E-mail: jaime.haro@upc.edu},
and 
Llibert Arest\'e Sal\'o$^{a,}$ \footnote{E-mail: llibert.areste@estudiant.upc.edu} 
}

\maketitle

\begin{center}
{\small

$^a$Departament de Matem\`atica Aplicada, Universitat
Polit\`ecnica de Catalunya \\ Diagonal 647, 08028 Barcelona, Spain \\

}
\end{center}

\thispagestyle{empty}

\thispagestyle{empty}

\begin{abstract}
We have tested some simple $\Lambda$CDM  (the same test is also valid for quintessence)
inflation models, imposing that they match with the recent observational data provided by the BICEP and Planck's team and leading to a reheating temperature, which is obtained via gravitational particle production after inflation, supporting the nucleosynthesis success
\end{abstract}

\vspace{0.5cm}



\maketitle





\section{Introduction}

There are several candidates to unify the early and late time acceleration of the universe, such as modified gravity \cite{nojiri1} or via phantom scalar fields \cite{nojiri2}.
 However, from our viewpoint 
 the best one to unify inflation and the current cosmic acceleration of the universe  is 
Quintessence or $\Lambda$CDM Inflation  { \cite{steinhardt, dv1, hossain,sami,berera}}. In this theory, the simplest way to construct the potential was performed for the first time in the seminal paper
\cite{pv} matching an inflationary one (used to
explain the early acceleration of the universe) with a quintessential one, which takes into account the current cosmic acceleration. For this kind of potentials,
at early times, the inflationary acceleration is the one that dominates and it ceases to be dominant in an abrupt phase transition. We note that {the phase transition needs to be abrupt in order to break the adiabatic regime, thus allowing a sufficient gravitational particle production . Moreover, an analytic expression of this amount 
 of particles is only obtained when the derivative of the potential presents some discontinuities \cite{ford,dv,gio,haro11} when  the universe enters in the kination regime \cite{Joyce}.  
The key point is that the energy density of the background decreases faster than the one of the produced particles, meaning that the energy density of the created particles 
 will eventually dominate, becoming the universe reheated and matching with the current hot Friedmann model.
  Finally, at very late times, the quintessential potential dominates and the universe starts to accelerate again. 

\

We point out that this kind of behavior could also be obtained by considering a universe with a small cosmological constant and by choosing a positive inflationary potential vanishing at some point, which is extended to zero for the other values of the field \cite{lh}. This comes from the fact that, when the potential is zero, the universe enters 
 in a kination regime which lasts until the energy density of the created particles at the phase transition starts to dominate, thus matching with the current $\Lambda$CDM model, in which the square root of the cosmological
 constant must be of the same order as the current value of the Hubble parameter in order to take into account the cosmic coincidence.

In order to assure the viability of these models, they are required to fit well with the recent observational data provided by the BICEP and Planck's teams \cite{bicep2, Planck}, but also the reheating temperature has to be compatible with nucleosynthesis, meaning that it is constrained approximately between $1$ MeV and $10^9$ GeV.
This is the 
 main goal of the present work, namely to study the viability of some well-known inflationary potentials adapted to 
quintessence or to $\Lambda$CDM model. 
Since our interest is to test the inflationary spectral parameters (spectral index, ratio of tensor to scalar perturbations, number of e-folds and reheating temperature of the universe), the quintessence piece of the potential plays no role in our calculations and, therefore, the two most important points to take into account are the inflationary phase and a phase transition to kination. For convenience, we choose
positive potentials that vanish at some point and we extend them to zero in order to ensure a phase transition to kination. Hence,
once we have these potentials, we calculate their spectral parameters, the number of e-folds { (as we will show, in quintessential or $\Lambda$CDM inflation it has to be between $63$ and $73$, which is a greater
range than the one given when the potential has a deep well)}
and its reheating temperature in two cases that will be analytically calculated:
via gravitational production of massless particles nearly conformally coupled to gravity, which is the most studied case in the literature,   
and very heavy massive particles conformally coupled to gravity.
 
\

The proceedings, which is essentially based on our recent paper \cite{lh}, is organized as follows: 
Section 2 is devoted to the study of the reheating via gravitational particle production. Two cases are studied in detail:  The gravitational production of heavy massive particles conformally coupled to gravity and 
the production of massless particles nearly conformally coupled to gravity. In Section 3 we calculate in two different ways the number of e-folds in quintessence inflation when there is a phase transition from inflation to kination, obtaining that it is constrainted to be between $63$ and $73$.
In Section 4, we consider a universe with a small cosmological constant and we adapt the simplest inflationary potentials appearing in the 
Encyclopaedia Inflationaris \cite{enciclopedia} to quintessence,
applying the results obtained in previous sections in order to study its viability.

\vskip 0.3cm

The units used throughout the paper are $\hbar=c=1$ and, with these units, $M_{pl}=\frac{1}{\sqrt{8\pi G}}$ is the reduced Planck's mass.

\section{Reheating in quintessence or $\Lambda$CDM inflation}

\subsection{Reheating via gravitational production of heavy massive particles conformally coupled to gravity}
In this section we will consider a pre-heating scenario that is not usually considered in quintessence: the creation of heavy massive particles conformally coupled to gravity that have no interaction with the inflaton field. In this situation,
the frequency of the particles in the $k$-mode is $\omega_k(\tau)=\sqrt{k^2+m_{\chi}^2a^2(\tau)}$, where $m_{\chi}$ is the mass of the quantum field.

During the adiabatic regimes, i.e., when
 $H(\tau)\ll m_{\chi} \Longrightarrow\omega_k'(\tau)\ll \omega_k^2(\tau)
  $, one can use the WKB approximation \cite{haro11}
\begin{eqnarray}
\chi_{n,k}^{WKB}(\tau)\equiv
\sqrt{\frac{1}{2W_{n,k}(\tau)}}e^{-{i}\int^{\tau}W_{n,k}(\eta)d\eta},
\end{eqnarray}
where $n$ is the order of the approximation of the $k$-vacuum mode. When some high order derivative
of the Hubble parameter is discontinuous, one has to match the $k$-vacuum modes before and after this moment. 
To do this, one needs to use positive
frequency modes after the breakdown of the adiabaticity, which is the cause of particle production. This is basically Parker's viewpoint of gravitational particle production \cite{parker}. Then, since the classical picture can be used at scales lower than the Planck's one, in order to preserve the condition $H(\tau)\ll m_{\chi} $ before the phase transition 
one has to choose
$m_{\chi}\geq M_{pl}$, but  elementary particles with masses greater than the Planck's one are micro black hole, whose behavior is unknown. Thus, to prevent the formation of these objects, we have to choose
$m_{\chi}\sim M_{pl}$ \cite{fkl}.

To clarify ideas and in order to obtain analytic expressions of the energy density of the produced particles, which allows us to calculate analytically the reheating temperature,  we consider a phase transition  from inflation to kination where the second derivative of the Hubble parameter is discontinuous.
If we assume that the derivative of the potential is discontinuous at $\varphi_E$, due to the conservation equation the second derivative of the inflaton field is discontinuous at the transition time and, consequently, from Raychaudhuri equation
$\dot{H}=-\frac{\dot{\varphi}^2}{2M_{pl}^2}$, one can deduce that the second derivative of the Hubble parameter is also discontinuous at the same time.

In this case one only needs the first order WKB solution to approximate the $k$-vacuum modes before and after the phase transition
\begin{eqnarray}
\chi_{1,k}^{WKB}(\tau)\equiv
\sqrt{\frac{1}{2W_{1,k}(\tau)}}e^{-{i}\int^{\tau}W_{1,k}(\eta)d\eta},
\end{eqnarray}
where
\begin{eqnarray}
W_{1,k}=
\omega_k-\frac{1}{4}\frac{\omega''_{k}}{\omega^2_{k}}+\frac{3}{8}\frac{(\omega'_{k})^2}{\omega^3_{k}} .
\end{eqnarray}

Before the transition time, namely $\tau=0$, vacuum is depicted by $\chi_{1,k}^{WKB}(\tau)$, but after the phase transition this mode becomes a mix of positive and negative frequencies of the form
$\alpha_k \chi_{1,k}^{WKB}(\tau)+\beta_k (\chi_{1,k}^{WKB})^*(\tau)$.

The $\beta_k$-Bogoliubov coefficient can be obtained matching both expressions at $\tau=0$, namely
\begin{eqnarray}
\beta_k=\frac{{\mathcal W}[\chi_{1,k}^{WKB}(t_E^-),\chi_{1,k}^{WKB}(t_E^+)]}
{{\mathcal W}[(\chi_{1,k}^{WKB})^*(t_E^+),\chi_{1,k}^{WKB}(t_E^+)]},
\end{eqnarray}
where ${\mathcal W}[f(t_E^-),g(t_E^+)]=f(t_E^-)g'(t_E^+)-f'(t_E^-)g(t_E^+)$ is the Wronskian of the functions $f$ and $g$ at the transition time, and $(+)$ (resp. $(-)$) means immediately after (resp. before) the phase transiton.

The square modulus of the $\beta$-Bogoliubov coefficient will be given approximately by
\begin{eqnarray}
 |\beta_k|^2\cong \frac{m^4a^{10}_E\left(\ddot{H}_E^+-\ddot{H}_E^-\right)^2}{256(k^2+m_{\chi}^2a^2_E)^5},
\end{eqnarray}
with
\begin{eqnarray}
\ddot{H}_E^+-\ddot{H}_E^-=-\frac{\dot{\varphi}_E}{M_{pl}^2}(\ddot{\varphi}_E^+- \ddot{\varphi}_E^-)
=-\frac{\dot{\varphi}_E}{M_{pl}^2}V_{\varphi}(t_E^-),
 \end{eqnarray}
 where we have used that in the kination phase the potential vanishes.

 This quantity, as we will see, is of the order $H_E ^3$ when dealing with quintessence models. Then 
 the number density of the produced particles and their energy density, {
 as has been rigorously proved in \cite{hll},}
will be
\begin{eqnarray}
 n_{\chi}(t)=\frac{1}{2\pi^2a^3}\int_0^{\infty}k^2|\beta_k|^2 dk\sim \bar\lambda^3\frac{H_E^{6}}{m_{\chi}^3}\left(\frac{a_E}{a(t)} \right)^3, \quad \rho_{\chi}(t)=\frac{1}{2\pi^2a^4}\int_0^{\infty}\omega_k(t)k^2|\beta_k|^2 dk\sim m_{\chi}n_{\chi}(t),
\end{eqnarray}
being $\bar\lambda$ a dimensionless constant, which depends on the model.

These massive particles will decay into lighter ones, which after some interactions become a fluid in thermal equilibrium.
To calculate the moment when this occurs, {
we assume that  the particles interact by exchange of gauge bosons and
we use the thermalization rate  $\Gamma=n_{\chi}(0)\sigma$ (see \cite{Allahverdi} and  also \cite{pv}), where the cross-section for emitting a gauge boson 
(whose typical energy is $E\sim \rho_{\chi}^{\frac{1}{4}}(0)$) from a scattering of two fermions is given by
 $\sigma\sim \frac{\alpha^3}{E^2}$,
being $\alpha$  a coupling constant with typical values $\alpha\sim 10^{-2}-10^{-1}$. Thus,  }
\begin{eqnarray}
 \Gamma=\alpha^3\left(\frac{n_{\chi}(0)}{m_{\chi}} \right)^{\frac{1}{2}}=\alpha^3\bar\lambda^{3/2}\frac{H_E^3}{m_{\chi}^2}.
\end{eqnarray}

Since  equilibrium is reached when $\Gamma\sim H(t_{eq})= H_E\left(\frac{a_E}{a_{eq}}\right)^3$,  we will have
\begin{eqnarray}
 \rho_{\chi}(t_{eq})\sim \alpha^3
 \bar\lambda^{9/2}\frac{H^8_E}{m_{\chi}^4}, \qquad
 \rho(t_{eq})\sim 3\alpha^6\bar\lambda^{3}\frac{H_E^6M_{pl}^2}{m_{\chi}^4},
\end{eqnarray}
and  the universe will become reheated when both energy densities are of the same order,
which will happen when $\frac{a_{eq}}{a_{R}}\sim \sqrt{\frac{\rho_{\chi}(t_{eq})}{\rho(t_{eq})}}$, {and so,  
\begin{eqnarray}
 T_R\sim \rho_{\chi}^{\frac{1}{4}}(t_R)\sim \rho_{\chi}^{\frac{1}{4}}(t_{eq})\sqrt{\frac{\rho_{\chi}(t_{eq})}{\rho(t_{eq})}} \sim
5\times 10^{-1} \alpha^{-3/4}
\bar \lambda^{15/8}\frac{H_E^3}{m_{\chi} M_{pl}}\sim 5\times 10^{-1} \alpha^{-3/4}
 \bar\lambda^{15/8}\frac{H_E^3}{M_{pl}^3} M_{pl},
\end{eqnarray}
where we have used that $m_{\chi}\sim M_{pl}$.

\subsection{Reheating via gravitational production of massless particles nearly conformally coupled to gravity}
In this situation, the Klein-Gordon equation is given by
\begin{eqnarray}\label{a2}
\bar{\chi}''_{k}(\tau)+\left(k^2+ \left(\xi-\frac{1}{6}\right)a^2(\tau)R(\tau)\right)\bar{\chi}_{ k}(\tau)=0,
\end{eqnarray}
where $\xi$ is the coupling constant and $R$ is the scalar curvature.

To define the vacuum modes before  and after the phase transition,
{ we use the in-out formalism (see \cite{starobinski},  \cite{birrell}, \cite{haro11}  or \cite{bd} for a review), where the behavior of these modes 
 at early and late times  is respectively}
\begin{eqnarray}
 \bar\chi_{b, k}(\tau)\simeq \frac{e^{-ik\tau}}{\sqrt{2{k}}} (\mbox{ when }\tau\rightarrow -\infty), \quad \bar\chi_{a, k}(\tau)\simeq \frac{e^{-ik\tau}}{\sqrt{2{k}}}
 (\mbox{ when } \tau\rightarrow +\infty).
\end{eqnarray}

Since we are considering particles nearly conformally coupled {to gravity}, we can consider the term $(\xi-1/6)a^2(\tau)R(\tau)$ as a perturbation, and we can approximate the ``b''
and ``a'' modes
by the first order Picard's iteration {as}
\begin{eqnarray}\label{a37}
 \bar\chi_{b, k}(\tau)\cong \frac{e^{-ik\tau}}{\sqrt{2{k}}}-\frac{\xi-1/6}{{k}\sqrt{2{k}}}
\int_{-\infty}^{\tau}a^2(\tau')R(\tau')\sin({ k}(\tau-\tau')) e^{-ik\tau'} d\tau',\\
\bar\chi_{a, k}(\tau)\cong \frac{e^{-ik\tau}}{\sqrt{2{k}}}+\frac{\xi-1/6}{{k}\sqrt{2{k}}}
\int_{\tau}^{\infty}a^2(\tau')R(\tau')\sin({ k}(\tau-\tau'))e^{-ik\tau'}d\tau',
\end{eqnarray}
which will represent, respectively, the vacuum before and after the phase transition.

Then, after the phase transition, we can write the ``in'' mode as a linear combination of the ``out'' mode and its conjugate as follows
\begin{eqnarray}
 \bar\chi_{b,k}(\tau)=\alpha_k\bar\chi_{a,k}(\tau)+\beta_k\bar\chi^*_{a,k}(\tau).
\end{eqnarray}

Imposing the continuity of $\bar\chi$ and its first derivative at the transition time we obtain,
up to order $\left(\xi-1/6  \right)^2 $,
that the value of these
coefficients is \cite{ford}
\begin{eqnarray}\label{bogoliubov}
  \alpha_k\cong 1-\frac{i({\xi}-\frac{1}{6})}{2k}\int_{-\infty}^{\infty}a^2(\tau)
 R(\tau) d\tau,\quad  \beta_k\cong \frac{i({\xi}-\frac{1}{6})}{2k}\int_{-\infty}^{\infty}e^{-2ik\tau}a^2(\tau)
 R(\tau) d\tau.
\end{eqnarray}

Finally, in order to define the energy density, { the in-out formalism is also used}, that is, when the universe is asymptotically static, the
energy density of the produced particles at late times is given by  $\frac{1}{2\pi^2a^4_{\infty}}\int_0^{\infty}k^3|\beta_k|^2 dk,$ where $a_{\infty}$ is the value of the scale factor at
late times. Then, despite not coinciding with the definition of energy density because the $1$-loop effects are disregarded (see for instance \cite{Glavan}  where the energy density for a transition from de Sitter to radiation is calculated),
it is assumed that
the energy density of the produced particles due to the phase transition is given by
\cite{bd}
\begin{eqnarray}
 \rho_{\chi}=\frac{1}{2\pi^2a^4}\int_0^{\infty}k^3|\beta_k|^2 dk.
\end{eqnarray}

The integral of the $\beta$-Bogoliubov coefficient (\ref{bogoliubov}) is convergent because at early and late times the term $a^2(\tau)R(\tau)$ converges fast enough to zero.
Moreover,
 if at the transition time $t_E$ the first derivative of the Hubble parameter is continuous, one has
$\beta_k\sim {\mathcal O}(k^{-3})$, which means that
 the energy density of the produced particles is not ultra-violet divergent. 
Then,
the energy density of the produced particles approximately becomes
\begin{eqnarray}
 \rho_{\chi}(t) \cong \left({\xi}-\frac{1}{6}\right)^2{\mathcal N}H^{4}_E\left(\frac{a_E}{a(t)}\right)^4,
\end{eqnarray}
where ${\mathcal N}$ is a dimensionless numerical factor and $H_E$ and $a_E$ are respectively the value of the Hubble parameter and the scale factor at the phase transition time.

\

{\bf Remark:}
The number ${\mathcal N}$ is clearly model dependent. In the case proposed by Ford in \cite{ford},  { the author considers a toy model} where there is
a transition from de Sitter to matter domination
modelled by $a^2(\tau)R(\tau)\equiv\frac{12}{\tau^2+\tau_0^2}$ and where $\mathcal  N$ can be analytically calculated giving as a result $\frac{9}{8}$. In \cite{dv}, a toy
model based on an abrupt transtion from de Sitter to radiation is considered, obtaining a result of the same order as Ford.
However,
note that in both cases reheating is impossible because the energy density of the produced particles decreases
faster or equal than those of the background. We have calculated numerically this factor for some simple models that have a transition from an inflationary regime to kination and in all
cases ${\mathcal N}$ is of the order 1.

\

 Finally, assuming that ${\mathcal N}\sim 1$, the reheating temperature, in this case, is given by
 \begin{eqnarray}\label{nearly}
 T_R\sim \left({\xi}-\frac{1}{6}\right)^{3/2}
 \left(\frac{H_E}{M_{pl}}\right)^2 M_{pl}.
 \end{eqnarray}

\section{Detailed calculation of the number of e-folds}

The number of e-folds can be calculated in two different ways:

\begin{enumerate}\item
By definition this quantity is equal to $N=\int_{t_*}^{t_{end}} H dt,$ where $(*)$ denotes when the pivot scale leaves the Hubble radius and $(end)$  stands for the end of  inflation. During the slow roll this quantity is given by
\begin{eqnarray}\label{slowroll}
N=\int_{t_*}^{t_{end}}Hdt\cong \frac{1}{M_{pl}^2}\int_{\varphi_*}^{\varphi_{end}}\frac{V}{V_{\varphi}}d\varphi.
\end{eqnarray}

\item
By using the whole history of the universe; in our case the transition from inflation to kination, passing though radiation and matter domination up to the present.  We start with the following equation \cite{liddle} 
\begin{eqnarray}
\frac{k_*}{a_0H_0}=e^{-N}\frac{H_*}{H_0}\frac{a_{end}}{a_E}\frac{a_E}{a_R}\frac{a_R}{a_M}\frac{a_M}{a_0}=e^{-N}\frac{H_*}{H_0}\frac{a_{end}}{a_E}\frac{\rho_R^{-1/12}\rho_M^{1/4}}{\rho_E^{1/6}}\frac{a_M}{a_0},
\end{eqnarray}
where $R$, $M$  and $0$ symbolize the beginning of radiation era, the beginning of the matter domination era and the present time, and having used relations $(a_E/a_R)^6=\rho_R/\rho_E$ (kination era) and $(a_R/a_M)^4=\rho_M/\rho_R$ (radiation domination). We use, as well, that $H_0\approx 2\times 10^{-4}\text{ Mpc}^{-1}$ and take as a physical value of the pivot scale $k_{phys}\equiv\frac{k_*}{a_0}=0.02\text{ Mpc}^{-1}$
(value used by Planck2015 \cite{Planck}). 
Moreover, we know that the process after reheating is adiabatic, i.e. $T_0=\frac{a_M}{a_0}T_M$, as well as the relations $\rho_M\approx\frac{\pi^2}{15}g_MT_M^4$ and $\rho_R\approx\frac{\pi^2}{30}g_RT_R^4$ (where $\{g_i\}_{i=R,M}$ are the relativistic degrees of freedom \cite{rehagen}). Hence,
\begin{eqnarray}
N=-4.61+\ln\left(\frac{H_*}{H_0} \right)+\ln\left(\frac{a_{end}}{a_E} \right)+\frac{1}{4}\ln\left(\frac{2g_M}{g_R}\right)+\frac{1}{6}\ln\left(\frac{\rho_R}{\rho_E} \right)+\ln\left(\frac{T_0}{T_R} \right).
\end{eqnarray}

We use that $H_0\sim 6\times 10^{-61}M_{pl}$ and, 
from the value of the power spectrum \cite{btw, blw} $P\approx \frac{H_*^2}{8\pi^2\epsilon_*M_{pl}^2}\sim 2\times 10^{-9}$,
we infere that $H_*\sim 4\times 10^{4}\sqrt{\epsilon_*}M_{pl}$, where $\epsilon_*$ is the main slow roll parameter evaluated when the pivot scale leaves the Hubble radius. We know as well that $T_0\sim 2\times 10^{-13} \text{ GeV}$ and $g_M=3.36$ \cite{rehagen}. Also, $g_R=107$, $90$ and $11$ for $T_R\geq 135$ GeV, $175 \text{ GeV}\geq T_R\geq 200$ MeV and $200\text{ MeV}\geq T_R\geq 1$ MeV, respectively \cite{rehagen}. On the other hand, assuming that the transition phase occurs immediately after the end of inflation and that there is not a substantial drop of energy, one obtains

\begin{eqnarray}
N\approx 54.5+\frac{1}{2}\ln\epsilon_*
-\frac{1}{3}\ln\left(\frac{g_R^{1/4}T_RH_{end}}{M_{pl}^2}\right). \label{eq13}
\end{eqnarray}

Therefore, with the values in our model and with the range $1\text{ MeV} \leq T_R\leq 10^9$ GeV required in order to have a successful nucleosynthesis \cite{giudice}, 
and taking, as usual, $H_{end}\sim 10^{-6} M_{pl}$ and $\epsilon_*\sim 10^{-2}$, we find that $63\leq N \leq 73$.  Moreover,  the equations (\ref{slowroll}) and (\ref{eq13}), as we will see, are functions
of the spectral index $n_s$. Then, equaling both equations, one will obtain some constraints for the spectral values in each model.

\end{enumerate}

\section{Application to some simple $\Lambda$CDM inflation {potentials}}

In this section we consider the simplest inflationary potentials that appear in \cite{enciclopedia} and, as we have explained in the introduction, we adapt them in order to have, after inflation,
a kination phase followed by the standard $\Lambda$CDM regime.
The way to obtain this kind of potentials is to choose, in the extensive list that appears in \cite{enciclopedia},
the simplest and most well-known positive inflationary potentials that vanish at some value of the scalar field and extend them to zero for the other values of the field. Moreover, in order to ensure the late time cosmic acceleration  and  coincidence,  we introduce a cosmological constant $\Lambda\sim H_0^2$, being $H_0$ the current value of the Hubble parameter.

\subsection{Exponential SUSY Inflation (ESI)}

The first potential we are going to study is  an Exponential SUSY Inflation (ESI) style potential  \cite{Obukhov, DT},
\begin{eqnarray}
V(\varphi)=\left\{ \begin{array}{ll} \lambda M_{pl}^4 (1-e^{\frac{\varphi}{M_{pl}}}) & \mbox{$\varphi<0$} \\ 0 & \mbox{$\varphi\geq 0$},  \end{array} \right.
\end{eqnarray}
being $\lambda$ a dimensionless positive parameter. By using the following approximate expressions of the slow-roll parameters as a function of the potential,
\begin{eqnarray}
\epsilon\approx\frac{M_{pl}^2}{2}\left(\frac{V_{\varphi}}{V}\right)^2 \ \ \ \eta\approx M_{pl}^2\frac{V_{\varphi\varphi}}{V},
\end{eqnarray}
we can compute the spectral index ($n_s$) and the ratio of tensor to scalar perturbations (r):
\begin{eqnarray}
n_s-1=-s_*\left(\frac{3}{2}s_*+2\right) \cong -2s_*
 \ \ \ \ r=16\epsilon_*, \label{eqpot1}
\end{eqnarray}
where we have introduced the notation $s_*=\frac{e^{\frac{\varphi_*}{M_{pl}}}}{1-e^{\frac{\varphi_*}{M_{pl}}}}\cong e^{\frac{\varphi_*}{M_{pl}}}$.
 It is also straightforward to calculate the power spectrum:
\begin{eqnarray}
P\approx \frac{H_*^2}{8\pi^2\epsilon_*M_{pl}^2}\approx \frac{\lambda}{3\pi^2 (1-n_s)^2},
\end{eqnarray}
where the constraint $P\sim 2\times 10^{-9}$ is verified by choosing $\lambda=6\times 10^{-9}\pi^2(1-n_s)^2$ . Finally, regarding the number of e-folds,
\begin{eqnarray}
N=e^{-\frac{\varphi_*}{M_{pl}}}-\frac{1+\sqrt{2}}{\sqrt{2}}+\frac{\varphi_*}{M_{pl}}-\ln\left(\frac{\sqrt{2}}{1+\sqrt{2}} \right)
\approx \frac{2}{1-n_s}+\ln\left(\frac{1-n_s}{2} \right).
\end{eqnarray}

Then,
in order to have $63\leq N\leq 73$ we need to take the range of $0.970\leq n_s\leq 0.974$ for the spectral index, obtaining thus $0.0013<r<0.0017$, which  matches with the observational 
Planck2015 data \cite{bicep2}.

On the other hand, to obtain reheating constraints we need to calculate $H_{end}$. Since inflation ends when $\epsilon_{end}=1$, i.e. when $\varphi_{end}=\ln\left(\frac{\sqrt{2}}{1+\sqrt{2}}\right)M_{pl}$,
using the definition {
 $\epsilon=-\frac{\dot{H}}{H^2}$} { \cite{btw}}, one has $\dot{H}_{end}=-H_{end}^2$ and, taking into account that 
$V=3H^2M_{pl}^2+\dot{H}M_{pl}^2$,
one can conclude that
$H_{end}=\sqrt{\frac{V(\varphi_{end})}{2M_{pl}^2} }=M_{pl}\sqrt{\frac{\lambda}{2(1+\sqrt{2})}}\approx 10^{-4}(1-n_s)M_{pl}$.

Therefore, by combining equation \eqref{eq13} and the just obtained approximate value of $N$, we have that
\begin{eqnarray}
Y-\ln Y=212.0-\ln\left(\frac{g_R^{1/4}T_R}{\text{GeV}} \right), \label{eq33}
\end{eqnarray}
where $Y\equiv\frac{6}{1-n_s}$.

Now, with regards to the case when the produced particles are very massive ( $m_{\chi}\sim M_{pl}$) and conformally coupled to gravity, we have to proceed as in Section $2.1$, obtaining the following  
energy density
$\rho_{\chi}\cong \frac{5H_{end}^6}{16^3\pi m^2}$, 
which leads to
\begin{eqnarray}
T_R\sim \frac{1}{\sqrt{3}}\left(\frac{5}{16\pi}\right)^{5/8}\left(\frac{H_{end}}{M_{pl}}\right)^2\frac{H_{end}}{m} M_{pl}\sim
10^{-1}\left( \frac{H_{end}}{M_{pl}}\right)^2\left(\frac{H_{end}}{m}\right) M_{pl} \sim 10^{-13}(1-n_s)^3M_{pl}.\end{eqnarray}

Finally, by combination with \eqref{eq33}, one gets
\begin{eqnarray}
Y-4\ln Y\cong 193.1,
\end{eqnarray}
obtaining that $n_s\cong 0.9720$, $r\cong 1.56\times 10^{-3}$ and $T_R\sim 5$ GeV. 

In order to study massless particles nearly conformally coupled to gravity,
we use the formula (\ref{nearly}) to get
\begin{eqnarray}
T_R\sim 10^{-8}\left|\xi-\frac{1}{6} \right|^{3/2}(1-n_s)^2 M_{pl},
\end{eqnarray}
obtaining, thus, that for the bounds of $T_R$ coming from the nucleosynthesis it should be verified that  $10^{-7}\leq\left|\xi-\frac{1}{6}\right|\leq 16$. This only gives us a restriction for the lower bound, since given that we have considered the particles to be nearly conformally coupled to gravity it should be verified that $\left|\xi-\frac{1}{6}\right|\leq 10^{-1}$, which adds the constraint that $T_R\leq 7\times 10^5$ GeV.

\subsection{Higgs Inflation (HI)}

Another potential that could work would be the following Higgs Inflation (HI) style potential in the Einstein Frame \cite{enciclopedia}.
\begin{eqnarray}
V(\varphi)=\left\{ \begin{array}{ll} \lambda M_{pl}^4 \left(1-e^{\frac{\varphi}{M_{pl}}}\right)^2, & \mbox{$\varphi<0$} \\ 0, & \mbox{$\varphi\geq 0$}  \end{array} \right.
\end{eqnarray}
being $\lambda$ a dimensionless positive parameter. In this case, 
$1-n_s\cong s_*$
being $s_*$ the same as above. In this case, the  number of e-folds is
\begin{eqnarray}
N=\frac{e^{-\frac{\varphi_*}{M_{pl}}}-(1+\sqrt{2})}{2}+\frac{\varphi_*}{2M_{pl}}-\frac{1}{2}\ln\left(\frac{1}{1+\sqrt{2}} \right)\sim \frac{1}{s_*}\cong \frac{1}{1-n_s}, 
\end{eqnarray}
which is bounded by $50$ for the allowed values of the spectral index. Therefore, the potential is not a viable quintessential inflation model because it leads to an insufficient number of e-folds.

\subsection{Power Law Inflation (PLI)}

Now, we are going to study a Power Law Inflation (PLI) \cite{Sahni} potential, adapted to quintessence
\begin{eqnarray}
V(\varphi)=\left\{ \begin{array}{ll} \lambda M_{pl}^4 \left(\frac{\varphi}{M_{pl}}\right)^{2n}, & \mbox{$\varphi<0$} \\ 0, & \mbox{$\varphi\geq 0$.}  \end{array} \right.
\end{eqnarray}

With the same procedure as in the former cases, we obtain 
$N=\frac{n+1}{1-n_s}-\frac{n}{2}$. Hence, we can easily verify that, so as to obtain a number of e-folds $63\leq N\leq 73$ with a spectral index $n_s=0.968\pm 0.006$, we need that $0.65<n<1.35$. 

\

Therefore, taking $n=1$ is a good choice in order to match our model with the observational results. Hence, if we consider the range of spectral index $0.9685\leq n_s\leq 0.973$, we obtain the desired number of e-folds. Regarding the ratio of tensor to scalar perturbations, which in our case is $r=4(1-n_s)$, the constraint $r\leq 0.12$ is verified for $n_s\geq 0.97$.
Finally, the power spectrum has the following expression
\begin{eqnarray}
P\approx \frac{4\lambda}{3\pi^2(1-n_s)^2}
\end{eqnarray}
and, thus, since $P\sim 2\times 10^{-9}$, we can determine $\lambda$, namely $\lambda\sim 10^{-11}$.

Now, we are going to proceed analogously as for the potential previously studied. Therefore, we approximate the Hubble constant at the transition point by $H_E\sim H_{end}=M_{pl}\sqrt{\lambda}\approx 10^{-4}(1-n_s)M_{pl}$. Now, combining Equation \eqref{eq13} and the number of e-folds for this particular potential, we obtain that

\begin{eqnarray}
Y+\frac{1}{2}\ln Y =213.5-\ln\left(\frac{g_R^{1/4}T_R}{\text{GeV}}\right), \label{eq47}
\end{eqnarray}
where, as before, $Y=\frac{6}{1-n_s}$.  From this equation we deduce that, in order to satisfy the constraint $r\leq 0.12$ { \cite{bicep2}}, the reheating temperature has to belong to $1$ MeV $\leq T_R\leq 10^4$ GeV.

\

Now, if we start considering the production of massless particles nearly conformally coupled to gravity, then the reheating temperature becomes
\begin{eqnarray}
T_R\sim \left|\xi-\frac{1}{6} \right|^{3/2}{\mathcal N}^{3/4}\frac{H_E^2}{M_{pl}}\sim 10^{-8}\left|\xi-\frac{1}{6}\right|^{3/2} (1-n_s)^2 M_{pl}.
\end{eqnarray}

Hence, for $1\text{ MeV} \leq T_R\leq 10^4$ GeV, we obtain that $10^{-7}\leq \left|\xi-\frac{1}{6}\right|\leq 5\times 10^{-3}$. On the other hand, in the case of heavy massive particles ($m_{\chi}\sim M_{pl}$), 
since the first derivative of the potential is continuous at the transition phase, one has to use for the $\beta$-Bogoliubov coefficient the expression given in \cite{inflation1},
\begin{eqnarray}
|\beta_k|^2\cong \frac{m^4 a_E^{12}(\dddot{H}_E^+-\dddot{H}_E^-)^2}{1024(k^2+m^2a_E^2)^6}.
\end{eqnarray}

Therefore, the energy density of the produced particles is equal to $\rho_{\chi}\cong \frac{10^{-3}}{\pi}\frac{H_{end}^8}{m^4}$. Thus,  
following step by step the calculations made in Section $2.1$,
 one gets the following reheating temperature
\begin{eqnarray}
T_R\sim5 \times 10^{-3}\left(\frac{H_{end}}{M_{pl}}\right)^2\left(\frac{H_{end}}{m}\right)^{9/4} M_{pl}\sim 5\times 10^{-16}(1-n_s)^{13/4} M_{pl}.
\end{eqnarray}

By combining it with equation \eqref{eq47}, namely 
\begin{eqnarray}
Y-\frac{11}{4}\ln Y\cong 200,
\end{eqnarray}
one obtains that $n_s\cong 0.9721$, $r\cong 0.1117$ and $T_R\sim 11$ MeV, which means that this potential supports the production of heavy massive particles.

\subsection{ Open String Tachionic Inflation  (OSTI)} 

We consider the following adapted form of the OSTI potential \cite{KL}
\begin{eqnarray}
V(\varphi)=\left\{ \begin{array}{ll} -\lambda M_{pl}^2 \varphi^2\ln\left[\left(\frac{\varphi}{\varphi_0}\right)^2\right], & \mbox{$\varphi< 0 $} \\ 0, & \mbox{$\varphi\geq 0$}  \end{array} \right.
\end{eqnarray}
where $|\varphi_0|\gg M_{pl}$.

For this potential one has $N\approx \frac{2}{1-n_s}-\frac{1}{2}$ and $r\approx 4(1-n_s)$. Thus, a number of e-folds comprised between $63$ and $73$ corresponds to a spectral index $0.9685<n_s<0.9728$ and a ratio of tensor to scalar perturbations $0.109<r<0.126$. Therefore, the constraint $r\leq 0.12$ restricts this range  to $66\leq N\leq 73$, corresponding to $0.97\leq n_s\leq 0.973$ and  $0.11<r<0.12$. As usual, the power spectrum $P\approx -\frac{4\lambda}{3\pi^2(1-n_s)^2}\ln\left[\left(\frac{M_{pl}}{\varphi_0}\right)^2\frac{8}{1-n_s} \right]$ is imposed to be $P\approx 2\times 10^{-9}$ by choosing the suitable value of $\lambda$.

\

Now, we approximate the Hubble constant at the transition point by $H_E\sim H_{end}=\sqrt{\frac{V(\varphi_{end})}{2M_{pl}^2}}\approx 10^{-4}\sqrt{1-n_s}M_{pl}$. The combination of equation \eqref{eq13} and the number of e-folds for this potential leads to
\begin{eqnarray}
Y+\ln Y=216.2-\ln\left(\frac{g_R^{1/4}T_R}{\text{GeV}}\right) \label{eq54}
\end{eqnarray}
obtaining, at $2\sigma$ C.L., that  the reheating temperature for this potential is constrained to be $1 \text{ MeV}\leq T_R\leq 10^5$ GeV.

As in the case of the power-law potential with $n=1$, this potential has a continuous derivative at the transition phase. Therefore, when considering the production of massless particles nearly conformally coupled to gravity, the reheating temperature is
\begin{eqnarray}
T_R\sim  10^{-8}\left|\xi-\frac{1}{6}\right|^{3/2}(1-n_s)M_{pl}.
\end{eqnarray}

So, for $1 \text{ MeV}\leq T_R\leq 10^5$ GeV, we obtain that $10^{-8}\leq \left|\xi-\frac{1}{6}\right|\leq 3\times 10^{-3}$.  However, dealing with heavy massive particles ($m_{\chi}\sim M_{pl}$), one can see that 
the second derivative of the potential, and thus the third derivative of the Hubble parameter, diverges at the transition phase, which means that we cannot use the WKB solution to approximate the
modes. Therefore, we are not able to compute, in this case, the reheating temperature.

\subsection{Witten-O'Raifeartaigh Inflation (WRI)}
 
In this case, the version of the WRI \cite{Witten, R} potential is
\begin{eqnarray}
V(\varphi)=\left\{ \begin{array}{ll} \lambda M_{pl}^4\ln^2\left(\frac{-\varphi}{|\varphi_E|} \right), & \mbox{$\varphi< -|\varphi_E| $} \\ 0, & \mbox{$\varphi\geq -|\varphi_E|$}  \end{array} \right.,
\end{eqnarray}
where $|\varphi_E|\gg M_{pl}$, obtaining that
\begin{eqnarray}
\varphi_*\approx -|\varphi_E|e^{\frac{4M_{pl}}{\sqrt{1-n_s}|\varphi_E|}} \ \ \ \ \varphi_{end}\approx -|\varphi_E|e^{\frac{\sqrt{2}M_{pl}}{|\varphi_E|}}.
\end{eqnarray}
Thus, 
with regards to the number of e-folds, it can be exactly integrated as $N=\frac{|\varphi_E|^2}{8M_{pl}^2}\left.\left(x^2(2\ln(x)-1)\right)\right|_{x_*}^{x_{end}}$, where we have introduced the notation  
$x\equiv\frac{-\varphi}{|\varphi_E|}$, whose expression up to order 2 in $\frac{M_{pl}}{|\varphi_E|}$ is $N\approx \frac{4}{1-n_s}-\frac{1}{2}\geq 100$, which results in a too high number of e-folds.

\subsection{K\"ahler Moduli Inflation I (KMII)}

The expression of the K\"ahler Moduli Inflation I (KMII) potential \cite{enciclopedia} is
\begin{eqnarray}
V(\varphi)=\left\{ \begin{array}{ll} \lambda M_{pl}^4\left(1-\alpha\frac{\varphi}{M_{pl}}e^{-\varphi/M_{pl}} \right), & \mbox{$\varphi>\varphi_E $} \\ 0, & \mbox{$\varphi\leq \varphi_E$}  \end{array} \right.,
\end{eqnarray}
where $\alpha$ is a positive dimensionless constant such that $\alpha\geq e$ and $\varphi_E$ is the value of $\varphi$ where $\frac{\varphi}{M_{pl}}e^{-\varphi/M_{pl}}=1/\alpha$ such that 
$\varphi_E\geq M_{pl}$. 

The number of e-folds can also be exactly integrated, namely
\begin{eqnarray}
N=x_{end}-x_*+\ln\left(\frac{x_{end}-1}{x_*-1}\right)+\frac{e}{\alpha}\left(Ei(x_*-1)-Ei(x_{end}-1) \right),
\end{eqnarray}
where $x=\varphi/M_{pl}$ and $Ei$ is the exponential integral function, which verifies that for $x_*\gg 1$, $Ei(x_*-1)\approx \frac{e^{x_*}}{x_*}$, being this the dominant term in the previous equation. Hence, by using that $x_*e^{-x_*}\approx\frac{n_s-1}{2\alpha}$, one obtains that
\begin{eqnarray}
N\approx \frac{2e}{1-n_s},
\end{eqnarray}
which does not fulfill our bounds for the number of e-folds and the spectral index. Therefore,   this potential is not a viable quintessential inflation model.

\subsection{Brane Inflation (BI)}

The Brane Inflation (BI) potential behaves as \cite{lmr}
\begin{eqnarray}
V(\varphi)=\left\{ \begin{array}{ll} \lambda M_{pl}^4\left[1-\left(\frac{-\varphi}{\mu M_{pl}}\right)^{-p} \right], & \mbox{$\varphi<\varphi_E$} \\ 0, & \mbox{$\varphi\geq \varphi_E$}  \end{array} \right.,
\end{eqnarray}
where $\mu$ and $p$ are positive dimensionless parameters and $\varphi_E\equiv -\mu M_{pl}$ . For simplicity  we only consider $p\in\mathbb{N}$ and we will consider two different cases:

\begin{itemize}

\item[a)] $\mu\ll 1$:

With the notation $x=\frac{-\varphi}{\mu M_{pl}}$, one will have 
$n_s-1\cong  -\frac{2p(p+1)}{\mu^2 x_*^{p+2}}$ and
\begin{eqnarray}
N=\frac{\mu^2}{p}\left[\frac{x_*^{p+2}}{p+2}-\frac{x_{end}^{p+2}}{p+2}-
\frac{x_*^2}{2}+\frac{x_{end}^2}{2}\right].
\end{eqnarray}

Taking into account that $x_{end}\cong \left( \frac{p^2}{2\mu^2} \right)^{\frac{1}{2(p+1)}}\ll x_*$, one has $N \cong \frac{\mu^2}{p}\frac{x_*^{p+2}}{p+2}$, meaning that
$N\approx \frac{2(p+1)}{(1-n_s)(p+2)}$, which enters in our range for values of $p$ greater than $17$.
For the tensor/scalar ratio one has $r\cong \frac{8p^2}{\mu^2 x_*^{2(p+1)}}\ll \frac{8p^2}{\mu^2 x_*^{p+2}}$ for $p\geq 1$, namely $r\ll \frac{4 p(1-n_s)}{p+1}\leq 4(1-n_s)$. Hence, one can conclude that for all the values of $2\sigma$ CL of the spectral index it is verified that $r<0.12$. And finally, by adjusting $\lambda$ so that $P\sim 2\times 10^{-9}$, we can build a successful quintessential model.

\

Effectively, regarding the reheating constraints, we obtain that for all the restricted values of the parameter $p$, $H_E\sim 10^{-4}\sqrt{1-n_s}M_{pl}$. So, as usual, one obtains that the reheating temperature bounds from nucleosynthesis give the constraint $0.968\leq n_s\leq 0.972$. For massive particles we have that $T_R\sim 10^3$ GeV. In the case of massless particles nearly conformally coupled to gravity, we obtain that $\left|\xi-\frac{1}{6}\right|\geq 10^{-8}$ and the fact that $\left|\xi-\frac{1}{6}\right|\leq 1$ should be satisfied constraints our reheating temperature to be less than $10^7$ GeV. 
\

\item[b)] $\mu\gg 1$:

In this case 
we have that
$1-n_s\cong 6\epsilon_*\cong \frac{1}{\mu^2}\frac{3p^2}{(1-x_*^p)^2}$. This means that $x_*\cong 1-\sqrt{\frac{3}{1-n_s}}\frac{1}{\mu}$, as well as 
$x_{end}\cong 1-\sqrt{\frac{1}{2}}\frac{1}{\mu}$. Consequently,
$N\sim \frac{1}{2}\left(\frac{3}{1-n_s} -1 \right)\sim 50$, which does not enter in our range.

\end{itemize}

\subsection{Loop Inflation (LI)}

In this case the potential behaves as \cite{Dvali}
\begin{eqnarray}
V(\varphi)=\left\{ \begin{array}{cc}
0 ,& \varphi\leq \varphi_E\equiv M_{pl}e^{-\frac{1}{\alpha}}\\
\lambda M_{pl}^4\left(1+\alpha \ln \left(\frac{\varphi}{M_{pl}}\right) \right),& \varphi\geq \varphi_E\equiv M_{pl}e^{-\frac{1}{\alpha}},
\end{array}\right.
\end{eqnarray}
where $\lambda$ and $\alpha$ are positive dimensionless constants. 

We consider two different asymptotic cases:
\begin{enumerate}
\item $0<\alpha \ll 1$:

In this case one has 
$n_s-1\cong \frac{2\alpha}{x^2_*}$, where we have introduced the parameter $x\equiv \frac{\varphi}{M_{pl}}$. For the number of e-folds one has
\begin{eqnarray}
N\cong \frac{x_*^2}{2\alpha}\cong \frac{1}{1-n_s},
\end{eqnarray}
which leads, as in the case of HI, to a not high enough number of e-folds.

 \item $\alpha \gg1$:

The spectral index and the tensor/scalar ratio will be as a function of $x_*$
\begin{eqnarray}
1-n_s=\frac{1}{x^2_*\ln^2 x_*}(3+2\ln x_*),\qquad r=\frac{8}{x^2_*\ln^2 x_*}.
\end{eqnarray}

Then, at $2\sigma$ C.L., for the allowed values of the spectral index, we can see, after some numerics, that $x_*$ ranges in the domain $6.94\leq x_*\leq 7.98$. On the other hand, the number of e-folds is 
\begin{eqnarray}\label{eq76}
N=\frac{x_*^2}{2}\left(\ln x_* -\frac{1}{2}\right)-\frac{x_{end}^2}{2}\left(\ln x_{end}-\frac{1}{2}\right).
\end{eqnarray}

Using the range of values for $x_*$ one finds that $34\leq N\leq 50$, which comes out of the viable range.

\end{enumerate}

\vspace{15cm}

\section{Discussion}

We have  adapted some inflationary potentials to { $\Lambda$CDM inflation}, extending them to zero after they vanish and adding a small cosmological constant. Once we have done it, we have tested the models imposing that: 
\

\begin{enumerate}
\item They fit well with the current observational data provided by BICEP and Planck teams. 

\

\item The number of e-folds  must range between $63$ and $73$. This number is larger than the usual one used for potentials with a deep well, due to the kination phase after inflation. 

\

\item The reheating temperature due to the gravitational particle production during the phase transition from inflation to kination has to be compatible with the nucleosynthesis success, i.e., it has to range between $1$ MeV and $10^9$ GeV.
\end{enumerate}

\
 
Our study shows that the potentials WRI and KMII lead to a too high number of e-folds, while for HI and LI potentials this number is too small. Other potentials such as ESI, PLI (only when the potential is quadratic), OSTI and BI satisfy the prescriptions $1$ and $2$.  Moreover, dealing with the reheating temperature, we have showed the viability of the reheating via the gravitational particle production of heavy massive particles conformally coupled to gravity and also via the production of massless particles nearly conformally coupled.

\vspace{6pt} 

\vspace{1cm}
{\bf Acknowledgments.}
{This investigation has
been supported in part by MINECO (Spain), Project 
No. MTM2014-52402-C3-1-P.
.}




\end{document}